\documentclass[12pt]{iopart}

\usepackage{latexsym,amssymb}
\usepackage{graphics}
\usepackage{graphicx}
\usepackage{epsfig}

\begin{document}

\title{Phase Transition in Two Species Zero-Range Process}

\author{M. R. Evans and T. Hanney}

\address{School of Physics and Astronomy, University of Edinburgh, Mayfield Road, Edinburgh, EH9 3JZ, UK}

\date{May 6, 2003}

\begin{abstract}
We study a zero-range process with two species of interacting
particles. We show that the steady state assumes a simple factorised
form, provided the dynamics satisfy certain conditions, which we
derive. The steady state exhibits a new mechanism of condensation
transition wherein one species induces the condensation of the other.
We study this mechanism for a specific choice of dynamics.
\end{abstract}

\pacs{05.70.Fh, 02.50.Ey, 64.60.-i}


\maketitle


\section{Introduction}

The zero-range process has recently received widespread attention in
the literature of nonequilibrium stochastic processes
\cite{spitzer,oloan,evans,kafri,kafri2,godreche,grosskinsky}. There
are three main reasons for this (see \cite{evans} for a review): (i)
it provides a minimal model for more complex systems, (ii) the steady
state has a simple factorised form, (iii) it exhibits a steady state
phase transition, between a fluid and a condensate phase. The simple
form of the steady state allows detailed analysis of this condensation
transition. Apart from being of interest in its own right (as an
example of a phase transition in an exactly solvable steady state),
the transition is of particular interest because of its relation to
coarsening phenomena in 1$d$ driven systems
\cite{oloan,kafri}. Further, it has provided a general criterion to
determine the existence of phase separation in driven systems with
conserved density \cite{kafri2}.

In the zero-range process, particles hop from site to site with hop
rates determined by the number of particles at the departure site.
Two types of condensation transition, whereby a finite fraction of the
particles condense onto a single site, have been studied \cite{evans}:
condensation induced by site-dependent hop rates and condensation
induced by the asymptotic dependence of the hop rate on particle number
at the site. The former is related to Bose-Einstein condensation
whereas  the latter involves a spontaneously broken symmetry.

Here, we generalise to a zero-range process with two species of
particles. The hop rates of each species are determined by the numbers
present of both species  at the departure site.
Under certain conditions which we specify, the steady state
again assumes a simple factorised form,
which we use to demonstrate a new mechanism of condensation
transition. In this case the condensation of one species
is induced by the distribution of particles of the other species.

One perspective on this two species model is as a system
of particles moving on an evolving landscape--- the dynamics of
one  particle species  depend locally on
how many particles of the other species are present (the landscape)
and these particles too are subject to locally determined
dynamics. The new condensation mechanism then is one where the
condensate is induced by the  landscape
whose evolution is in turn coupled to the particle distribution. 
Such interplay between particle
dynamics  and an evolving landscape is  
of considerable interest \cite{drossel,das}.

\section{Steady State} 
\label{SECT:SS}
  
We define a two species zero-range process as follows.
Consider a chain of $L$ sites, with periodic boundary conditions, and
containing $N$ particles of species $A$ and $M$ particles of species
$B$. Sites, labelled $l=1,\ldots,L$,  contain $n_l$ particles of
species $A$ and $m_l$ particles of species $B$. Particles hop to their
nearest neighbour site to the right, species $A$ with rate $u(n,m)$
and species $B$ with rate $v(n,m)$, where $n$ and $m$ are the occupation
variables of the departure site. Thus the hopping rates, which depend
on the number of particles of both species at the departure site,
contain the interaction between species.

The steady state for the two species model can be derived in a similar
way to that for the single species zero-range process \cite{evans}.
First, we define $P(\{n_l\};\{m_l\})$ to be the steady state
probability of finding the system in the configuration
$(\{n_l\};\{m_l\})$, where $\{n_l\}=n_1,\ldots,n_L$ and
$\{m_l\}=m_1,\ldots,m_L$. Now, the steady state Master equation, which
demands that the probability current into a particular configuration
is equal to the probability current out of the same configuration, can
be written 
\begin{eqnarray} \label{ME}
\fl 0 = \sum_{l=1}^L \left[ \left\{ u(n_{l-1}\!+\!1,m_{l-1})
P(\ldots,n_{l-1}\!+\!1,n_l\!-\!1,\ldots;\ldots,m_{l-1},m_l,\ldots)
\right. \right. \nonumber \\
\left. - u(n_l,m_l) P(\ldots,n_{l-1},n_l,\ldots;\ldots,m_{l-1},m_l,\ldots)
\right\} \theta(n_l) \nonumber \\ 
+ \left\{ v(n_{l-1},m_{l-1}\!+\!1)
P(\ldots,n_{l-1},n_l,\ldots;\ldots,m_{l-1}\!+\!1,m_l\!-\!1,\ldots)
\right. \nonumber \\ 
\left. \left. - v(n_l,m_l)
P(\ldots,n_{l-1},n_l,\ldots;\ldots,m_{l-1},m_l,\ldots) \right\}
\theta(m_l) \right]
\end{eqnarray}
where $\theta(x)$ is the usual Heaviside function and ensures that the
site $l$ is occupied in order for a particle either to have arrived or
to be able to vacate there. The first term on the r.h.s of (\ref{ME})
is a gain term due to an $A$ particle hopping into site $l$
from site $l\!-\!1$, the second term is a loss term due to an $A$ particle
hopping out of site $l$; the third and fourth terms represent
analogous processes for the $B$ particles. We seek a solution to (\ref{ME})
where $P(\{n_l\};\{m_l\})$ is given by a product measure (i.e. a
product of factors)  
\begin{equation} \label{PM}
P(\{n_l\};\{m_l\}) = Z_{L,N,M}^{-1} \prod_{l=1}^{L} f(n_l,m_l) \;,
\end{equation}
where $Z_{L,N,M}$ is a normalisation. 
Equation (\ref{ME}) can be satisfied by substituting in the steady
state (\ref{PM}), and asking that the gain and loss
terms due to the dynamics
of the $A$ particles cancel independently of the gain and loss terms
due to the dynamics of the
$B$ particles. Then, if this cancellation is achieved for each term
$l$ in the sum individually, one finds that for all $l$,
\begin{equation}
\fl u(n_l,m_l) f(n_{l-1},m_{l-1})
f(n_l,m_l) = u(n_{l-1}\!+\!1,m_{l-1})f(n_{l-1}\!+\!1,m_{l-1})
f(n_l\!-\!1,m_l) \;,
\end{equation}
for all $n_l \neq 0$, and  
\begin{equation}
\fl v(n_l,m_l) f(n_{l-1},m_{l-1})
f(n_l,m_l) = v(n_{l-1},m_{l-1}\!+\!1)f(n_{l-1},m_{l-1}\!+\!1)
f(n_l,m_l\!-\!1) \;,
\end{equation} 
for all $m_l \neq 0$. These equations in turn imply that 
\begin{eqnarray} \label{cancelAs}
\frac{u(n_l,m_l) f(n_l,m_l)}{f(n_l\!-\!1,m_l)} =
\frac{u(n_{l-1}\!+\!1,m_{l-1})f(n_{l-1}\!+\!1,m_{l-1})}{f(n_{l-1},m_{l-1})}
= {\rm constant} \;, \\
\label{cancelBs}
\frac{v(n_l,m_l) 
f(n_l,m_l)}{f(n_l,m_l\!-\!1)} =
\frac{v(n_{l-1},m_{l-1}\!+\!1)f(n_{l-1},m_{l-1}\!+\!1)}{f(n_{l-1},m_{l-1})}
= {\rm constant} \;,
\end{eqnarray}
for all $l$. Both constants are set equal to unity (they only appear as
an overall factor in the normalisation). The two relations
(\ref{cancelAs}) and (\ref{cancelBs}) imply a constraint on the choice
of $u(n_l,m_l)$ and $v(n_l,m_l)$ i.e. we can use
(\ref{cancelAs}) and (\ref{cancelBs}) to obtain two expressions for
$f(n_l,m_l)$ in terms of $f(n_l\!-\!1,m_l\!-\!1)$:
\begin{equation}
f(n_l,m_l) = \frac{f(n_l\!-\!1,m_l\!-\!1)}{u(n_l,m_l) v(n_l\!-\!1,m_l)}
= \frac{f(n_l\!-\!1,m_l\!-\!1)}{u(n_l,m_l\!-\!1) v(n_l,m_l)}
\end{equation}
but both of these expressions must give the same result, therefore the
hopping rates are required to obey the constraint,
\begin{equation} \label{CE}
\frac{u(n_l,m_l)}{u(n_l,m_l\!-\!1)} = \frac{v(n_l,m_l)}{v(n_l\!-\!1,m_l)} \;,
\end{equation}
for $n_l,m_l \neq 0$. The choices of $u(n_l,0)$ and $v(0,m_l)$ remain
unconstrained.  

Finally, the expression for $f(n_l,m_l)$ is obtained by iterating
(\ref{cancelAs}) and (\ref{cancelBs}), hence  
\begin{equation} \label{fnm}
f(n_l,m_l) = \prod_{i=1}^{n_l} \left[ u(i,m_l) \right]^{-1}
\prod_{j=1}^{m_l} \left[ v(0,j) \right]^{-1} \;,
\end{equation}
where we have set $f(0,0) = 1$. Although not immediately obvious,
$u(n,m)$ and $v(n,m)$ do play symmetric roles in
(\ref{fnm}), but this symmetry is obscured within the constraint (\ref{CE})
on the hopping rates. 

Also note that we have the freedom to
choose \emph{any} desired form for $f(n_l,m_l)$, 
then by substituting this form
into (\ref{cancelAs}) and (\ref{cancelBs}) we deduce the hopping rates
which would lead to such a steady state---such rates are then
guaranteed to satisfy the condition (\ref{CE}).


In order to look for phase transitions, it will prove useful to consider the
normalisation, $Z_{L,N,M}$, defined in equation (\ref{PM}). This
quantity plays a role analogous to the canonical partition
function of equilibrium statistical mechanics. It can be written 
\begin{equation} \label{zlnm}
Z_{L,N,M} = \sum_{\left\{ n_l \right\} , \left\{ m_l \right\}}
\delta( \sum_{l=1}^L n_l - N) \delta( \sum_{l=1}^L m_l - M)
\prod_{l=1}^{L} f(n_l,m_l) \;,
\end{equation}
where the delta-functions ensure that we have $N$ particles of species
$A$ and $M$ particles of species $B$ in the system. 
Using an integral representation of the
delta-functions  yields
\begin{equation} \label{NORM}
Z_{L,N,M} = \sum_{\left\{ n_l \right\} , \left\{ m_l \right\}}
\oint \frac{dz}{2 \pi i} \oint \frac{dy}{2 \pi i}
\frac{\prod_{l=1}^{L} f(n_l,m_l)z^{n_l}y^{m_l}}{z^{N+1}y^{M+1}} \;,
\end{equation}
which leads us naturally to define the generating function
\begin{equation} \label{FZY}
F(z,y) = \sum_{n=0}^{N} \sum_{m=0}^M z^n y^m f(n,m)\;.
\end{equation}

We now assume that in the limit $N$, $M$, $L\to\infty$ (with the
particle densities held fixed), the integral
in equation (\ref{NORM}) is dominated by its saddle point. The saddle
point equations, which involve the particle densities of species
$A$ and $B$, $\rho_A=N/L$ and $\rho_B=M/L$, are found to be 
\begin{equation} \label{spd}
\rho_A = z \frac{\partial}{\partial z} {\rm ln} F(z,y) \;, \qquad
\rho_B = y \frac{\partial}{\partial y} {\rm ln} F(z,y) \;.
\end{equation}
These equations determine $z$ and $y$. For the saddle point to be
valid, $z$ and $y$ must be less than or equal to the radii of
convergence of $F(z,y)$ in order that we can perform the sum
(\ref{FZY}) in the first place. We note also
that because all derivatives of $F(z,y)$ are positive, the saddle
point, if valid, must be unique. In the following, we will find that
it is only possible to solve (\ref{spd}) in the
allowed ranges of $z$ and $y$ for certain values of $\rho_A$ and
$\rho_B$.  

\section{Phase Transition---analysis} 
\label{SECT:PT}

We now consider how a condensation transition may arise from the
interaction of the two species.
A general case of interest is where the hopping rates of one of the
particle species depend only on the number of the particles of the
other species at the site e.g. $v(n,m) = r(n)$. Then, by (\ref{cancelBs}),
$f(n,m) = r(n)^{-m} s(n)$, and $u(n,m)$ is determined by
(\ref{cancelAs}) ($r(n)$ and $s(n)$ are general functions of $n$). In
this case, the $A$ particles play the role of an evolving landscape
that determines the dynamics of the $B$ particles. For 
simplicity, we consider the case where $r(n) = 1+c/(n+1)$ and $s(n)=1$, then
\begin{equation}
v(n,m)=1+\frac{c}{n+1}\,,  \qquad
u(n,m)=\left( \frac{1+\frac{c}{n+1}}{1+\frac{c}{n}} \right)^m\,,
\label{rates}
\end{equation}
for all values of $n$ and $m$; $c$ is a constant. Physically, these
hopping rates are such that the $B$ particles hop more slowly as the
number of $A$ particles at the departure site increases, while the $A$
particles slow as the number of $B$ particles increases but they hop
more quickly the more particles of the same species occupy the
departure site. 
The generating function and its first derivatives, having performed
the sum over $m$, are obtained as 
\begin{eqnarray}
\label{fzy}
F(z,y) = \sum_{n=0}^{\infty} z^n \frac{1+c+n}{(1+n)(1-y)+c}\,,\\
\label{zdzfzy}
z \frac{\partial}{\partial z} F(z,y) = \sum_{n=0}^{\infty} n z^n
\frac{1+c+n}{(1+n)(1-y)+c}\,, \\
\label{ydyfzy}
y \frac{\partial}{\partial y} F(z,y) = \sum_{n=0}^{\infty} (1+n) z^n
\frac{y(1+c+n)}{[(1+n)(1-y)+c]^2}\,.
\end{eqnarray}
The radii of convergence of these series are $z=1$ and
$y=1$. These expressions determine $z$ and $y$, given $\rho_A$ and
$\rho_B$, as prescribed by the saddle point equations (\ref{spd}). 

\begin{figure} 
\begin{center}
\includegraphics[scale=0.5]{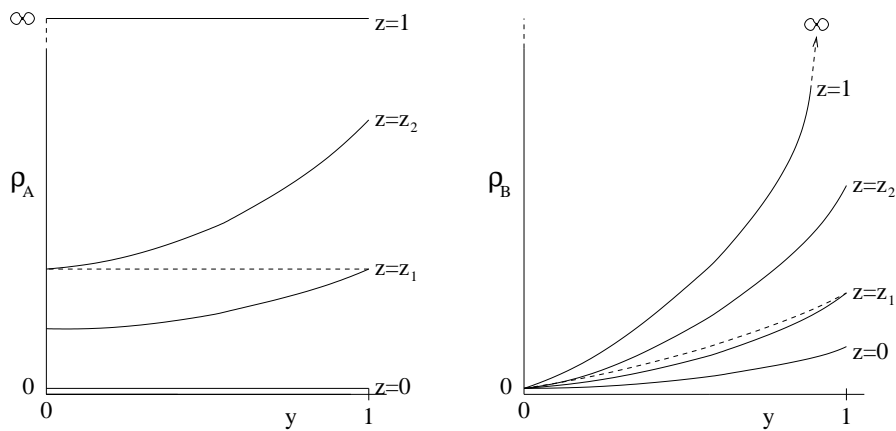}
\end{center} 
\caption{Schematic dependences of the particle densities $\rho_A$
and $\rho_B$ for 
contours of fixed $z$ and as a function of $y$. The dashed line
in the right hand graph
illustrates how $\rho_B$ varies as a function of $z$ and $y$ given
that $\rho_A$ is fixed.} 
\label{fig:dense}
\end{figure}

\begin{figure}
\begin{center}
\unitlength=1mm
\linethickness{0.4pt}
\begin{picture}(55,55)

\put(5,5){\line(1,0){50}}
\put(5,5){\line(0,1){50}}
\put(20.2,5.2){\line(1,1){32}}

\put(14,33){Fluid phase}
\put(35,16){Condensate}
\put(40,12){phase}

\put(0.5,45){$\rho_A$}
\put(45,2){$\rho_B$}

\put(17,1.6){$1/c$}

\end{picture}
\caption{Phase diagram for hopping rates (\ref{rates}).} \label{fig:phase_diagram}
\end{center}
\end{figure}
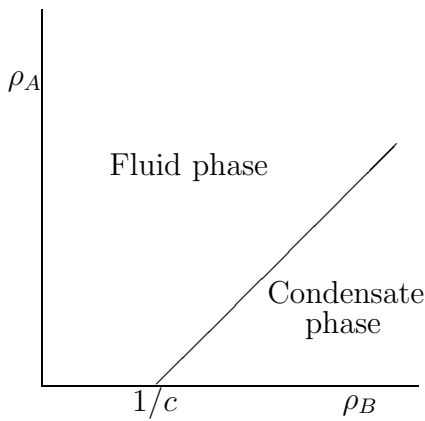

One can demonstrate the existence of a phase transition in the
following way.
For a given $z$, both $\rho_A$ and $\rho_B$ are
monotonically increasing functions of $y$. Then by analysing the
saddle point equations for the four cases $y=0$, $z=0$, $y=1$, and $z=1$, the
dependences of $\rho_A$ and $\rho_B$ on $y$, for fixed
$z$, are obtained. These are shown in figure \ref{fig:dense}, where $0
< z_1 < z_2 < 1$.
Now, if we consider a system containing a density $\rho_A$ of $A$
particles, then the solution of (\ref{spd}) requires that $z$ and $y$
lie in the range 
$z_1 \leq z \leq z_2$ and $0 \leq y \leq 1$. In this range, $\rho_B$
increases monotonically from $\rho_B=0$, where $z=z_2$ and $y=0$, to
its maximum value allowed by the saddle point equations, where $y=1$
and $z=z_1$. 
Therefore we can only solve the saddle point equations for $\rho_A$
and $\rho_B$ if $\rho_B$ is less than this ($\rho_A$-dependent)
maximum. 
When $\rho_B$ exceeds this maximum,
the saddle point approximation is no longer valid and a 
phase transition ensues. The critical line is given by the 
values of $\rho_A$ and $\rho_B$, as a function of $z$, where
$y=1$. Along this line $\rho_A$ and
$\rho_B$ are related by 
\begin{equation} \label{critline}
\rho_B = (1+\rho_A)/c \,.
\end{equation}
This is most easily seen by setting $y=1$ in (\ref{fzy}),
(\ref{zdzfzy}) and (\ref{ydyfzy}), and then spotting that one can write $y
\partial F(z,y) / \partial y = [F(z,y) + z \partial F(z,y) / \partial
z]/c$. Then dividing through by $F(z,y)$ yields (\ref{critline}). 
Hence we obtain the phase diagram shown in figure \ref{fig:phase_diagram}. 

\begin{figure} 
\begin{center}
\includegraphics[scale=0.27,angle=270]{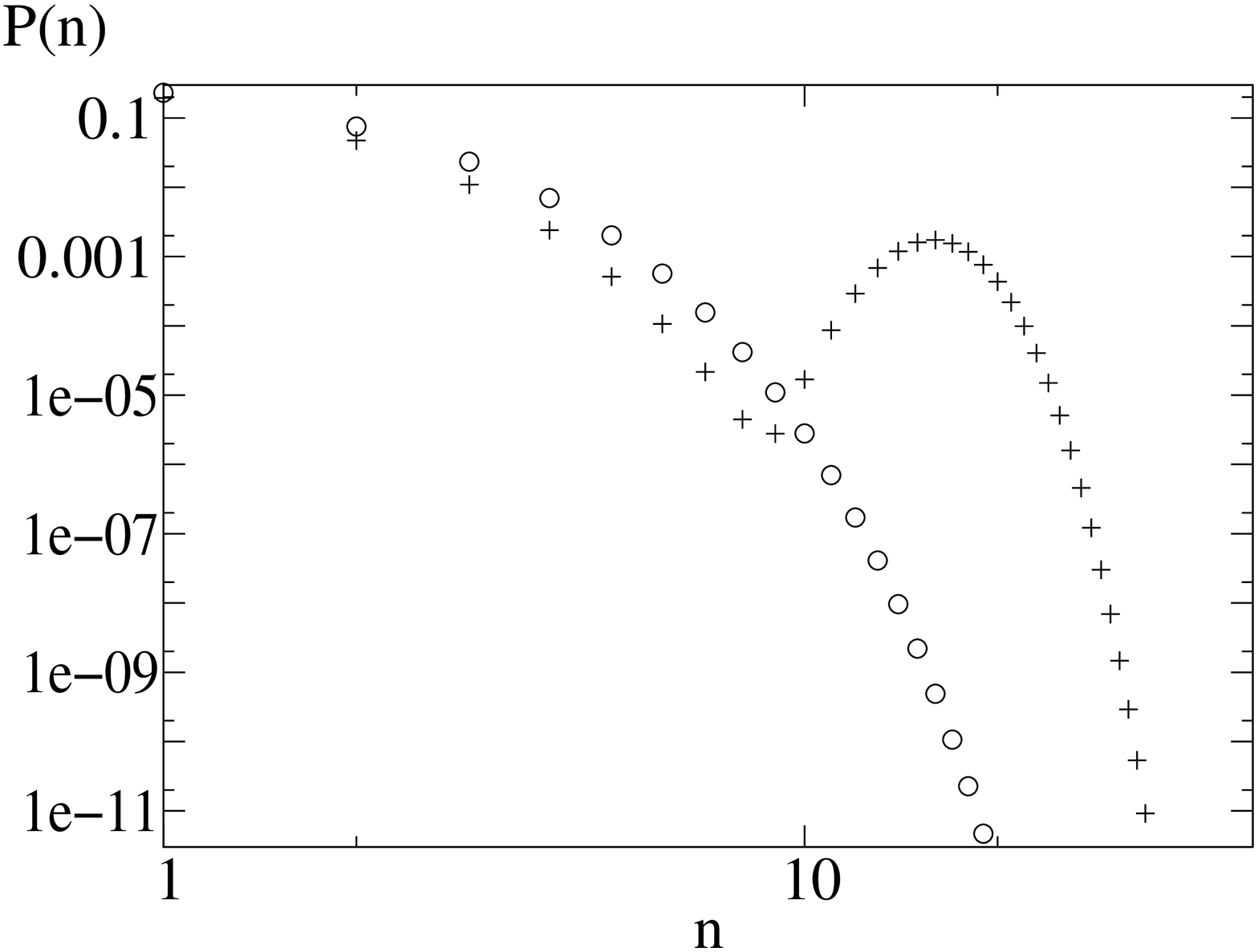}
\includegraphics[scale=0.27,angle=270]{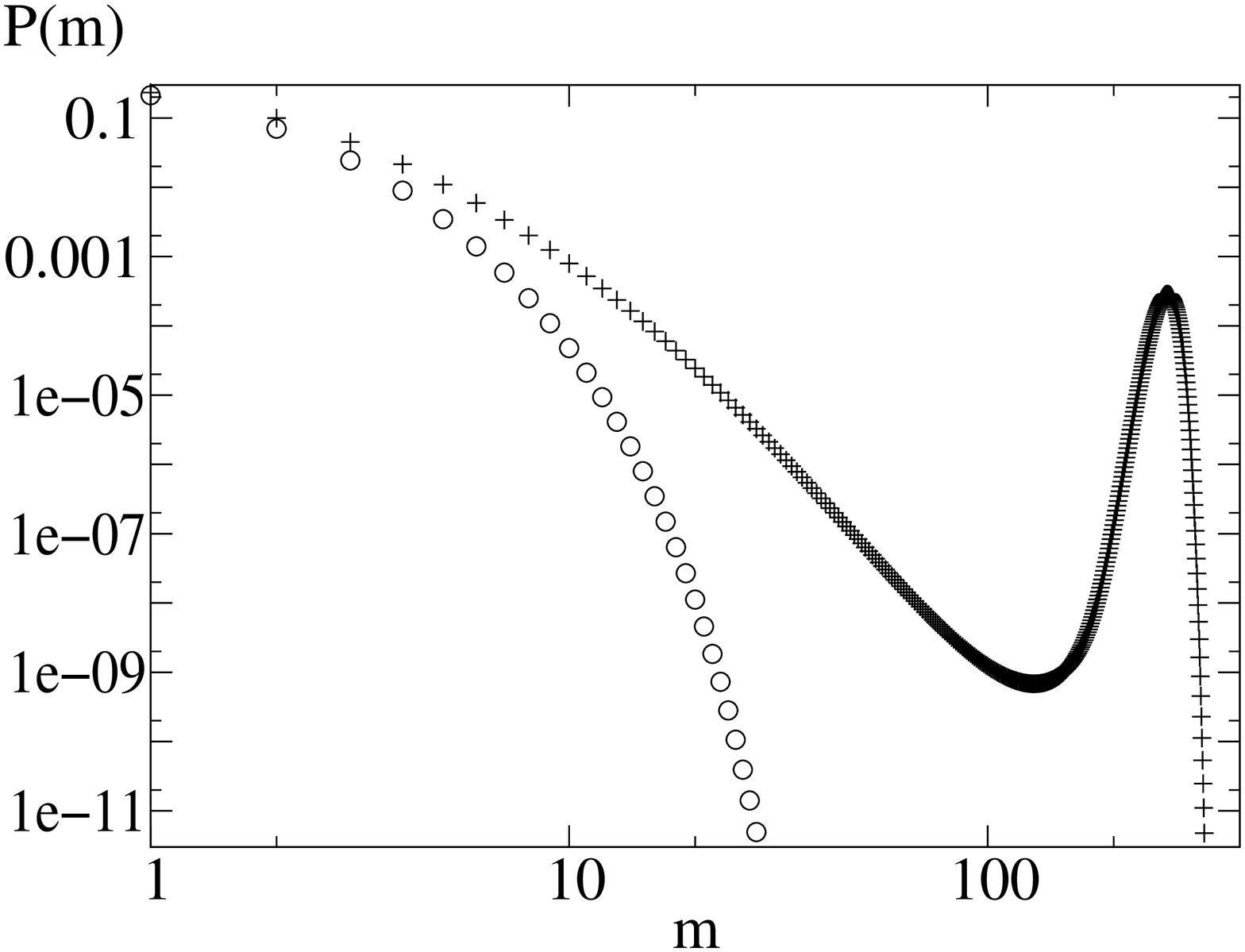}
\end{center} 
\caption{Log-log plot of on the left $P(n)$ vs. $n$, and on the right
$P(m)$ vs. $m$, for systems of size $L=100$ and $c=2$. The circles
correspond to densities $\rho_A = 1/2 = \rho_B$; the crosses
correspond to densities $\rho_A = 1/2$ and $\rho_B = 7$.} 
\label{fig:dists}
\end{figure}

We interpret this as a transition between a fluid phase, where the
saddle point equations can be solved, and a condensate phase. In the
condensate phase, $\rho_B$ exceeds a
maximum critical value and the excess $B$ particles condense onto a
single site. The number of $B$ particles in the condensate is
proportional to the system size $L$---in the limit $N$, $M$, $L\to
\infty$ it contains a finite fraction of an
infinite number of particles. However, we note that
for $m \to \infty$, $u(n,m) \to 0$ if $n$ is finite. But the current
of $A$ particles must be the same across all bonds in the system as a
result of continuity, and this current must be finite---the current of
$A$ particles away from the condensate is finite---therefore we
must have that $n \to \infty$ at the condensate in order that the
current out of the condensate site is finite. For $n$ large, $u(n,m)
\sim {\rm exp}(m/n^2)$ and this must be finite, therefore $m \sim n^2$
at the condensate then since $m \propto L$ here, $n \propto L^{1/2}$. 
Thus the condensate phase is characterised by a condensate of $B$
particles which is sustained by a `weak' condensate of $A$ particles.
This picture is consistent with numerical results: an exact
recursion relation for the partition function can be obtained from
(\ref{zlnm}) and is found to be 
\begin{equation}
Z_{L,N,M} = \sum_{n=0}^N \sum_{m=0}^M f(n,m) Z_{L-1,N-n,M-m} \,.
\end{equation}
This expression was iterated exactly on a computer up to systems of
size $L=100$, for different values of the densities and using $c=2$. It is also
straightforward to obtain an exact expression for $P(n,m)$, the
probability that a site contains exactly $n$ particles of species $A$
and $m$ particles of species $B$. $P(n,m)$ is given by
\begin{equation}
P(n,m) = Z_{L,N,M}^{-1} Z_{L-1,N-n,M-m} f(n,m) \,.
\end{equation}
By summing this equation over $n$ or $m$ one obtains equations for
$P(n)$, the probability of finding exactly $n$ particles of species
$A$ at a site, and $P(m)$, the probability of finding exactly $m$
particles of species $B$ at a site. These probabilities are plotted in
figure \ref{fig:dists} for two different values of the densities. 
The circles correspond to densities $\rho_A = 1/2 = \rho_B$ and the
system is in the fluid phase---both $P(n)$ and $P(m)$, as expected,
decay exponentially. For $\rho_A = 1/2$ and $\rho_B = 7$,
represented by the crosses, the system is in the condensate phase. 
Within finite size effects, these 
results are consistent with the following 
picture: the condensate of $B$
particles (containing $\sim L$ particles) is sustained by a `weak'
condensate of $A$ particles (containing $\sim L^{1/2}$
particles); the condensates exist on
a power-law distributed background of $B$
particles forming a  critical fluid,
and an exponentially distributed
background of $A$ particles forming a normal fluid.
 
\section{Conclusion}
\label{CONC}

We have shown that one may define a two species zero-range process
where the steady state is described by a product measure (\ref{PM})
and we have derived conditions under which the product measure holds
(\ref{CE}). This simple form for the steady state makes the model
amenable to detailed analysis. In particular, we have shown that
the model undergoes a novel kind of condensation transition.
  
It is clearly of interest to investigate the generality of this
transition with respect to the hopping rates.  Since we can always
construct dynamics, using (\ref{cancelAs}, \ref{cancelBs}), that lead
to a product measure (\ref{PM}) with arbitrary $f(n,m)$ there is scope
within the model for further mechanisms of condensation
transition \cite{HE}. Further analysis of the nature of the weak condensate is
also desirable.

\ack
We thank David Mukamel for useful discussions. TH thanks EPSRC for financial
support under grant GR/52497.

\end{document}